\documentclass{PoS}

\title{Pions: Experimental Tests of Chiral Symmetry Breaking}

\ShortTitle{AB-Overview}

\author{\speaker{A.M. Bernstein}\thanks{I would like to thank my colleagues in the A2 experiments at Mainz, particularly D. Hornidge and S. Schumann, as well as C. Fernandez-Ramirez for our productive collaboration. For sharing information about their work and providing figures I would like to thank J. Camalich, S. Durr, J. Friedrich, J. Pelaez, D. Lawrence, and D. Renner. For stimulating discussions I would like to thank H. Leutwyler, B. Holstein, and  A. Walker-Loud. For helpful comments on the manuscript I thank P. Martel and J. Goity. My research is supported in part by a DOE grant to the Laboratory for Nuclear Science, MIT.}
\\
       Physics Dept. and Laboratory for Nuclear Science\\
         Massachusetts Institute of Technology \\ 
         Cambridge MA. 02139, USA\\     
        E-mail: \email{bernstein@mit.edu}}

\abstract{
Based on the spontaneous breaking of chiral symmetry, chiral perturbation theory (ChPT) is believed to approximate confinement scale QCD. Dedicated and increasingly accurate experiments and improving lattice calculations are confirming this belief, and we are entering a new era in which we can test confinement scale QCD in some well chosen reactions. This is demonstrated with an
overview of low energy experimental tests of ChPT predictions of $\pi\pi$ scattering, pion properties, $\pi$N scattering and electromagnetic pion production. These predictions have been shown to be consistent with QCD in  the meson sector by increasingly accurate lattice calculations. At present there is good agreement between experiment and ChPT calculations, including the $\pi\pi$ and $\pi$N s wave scattering lengths and the $\pi^{0}$ lifetime. Recent, accurate pionic atom data are in agreement with chiral calculations once isospin breaking effects due to the mass difference of the up and down quarks are taken into account, as was required to extract the $\pi\pi$ scattering lengths. In addition to tests of the theory, comparisons between $\pi\pi$ and $\pi$N interactions based on general chiral principles are discussed.  Lattice calculations are now providing results for the fundamental, long and inconclusively studied, $\pi$N $\sigma$ term and the contribution of the strange quark to the mass of the proton. Increasingly accurate experiments in electromagnetic pion production experiments from the proton which test ChPT calculations (and their energy region of validity) are presented. These experiments are also beginning to measure the final state $\pi$N interaction. This paper is based on the concluding remarks made at the Chiral Dynamics Workshop CD12 held at Jefferson Lab in Aug. 2012.}

\FullConference{The 7th International Workshop on Chiral Dynamics,\\
		August 6 -10, 2012\\
		Jefferson Lab, Newport News, Virginia, USA}

\begin{document}

\section{Chiral Symmetry Breaking in QCD}
As was first realized by Nambu, chiral symmetry is spontaneously broken (hidden)\cite{Nambu}. In the limit of massless light quarks (u,d) this leads to massless, pseudoscalar, pions (see e.g.\cite{Nambu,ChPT,book}). The small, explicit, chiral symmetry breaking due to the finite light quark masses gives the pions a non-zero mass. At low energies quarks and gluons are confined and QCD is a non-linear theory which can be solved numerically by lattice techniques. It can be approximated at low energies by a systematic expansion in the momenta and masses of the hadrons (pions, nucleons, ...) by an effective field theory known as Chiral Perturbation Theory (ChPT)\cite{ChPT,book,ChPT-review}, which has been tested with increasingly accurate experiments and more recently with lattice calculations. These tests are the primary subject of this paper.
In addition to leading to precise tests, the hiding of chiral symmetry means that the interactions of pions with other hadrons at low energies is weak in the s wave close to threshold and strong in the p wave. This leads to some general properties for pions and nucleons which are of interest. Some of these will be explored in this paper. Tests of pion properties such as the $\pi^{0}$ lifetime and $\pi\pi$ scattering have reached new levels of accuracy, both in experiment and chiral and lattice calculations. Success in the more difficult $\pi$N sector are somewhat behind but significant progress is being made\cite{ChPT-Baryons}[Meissner]\footnote{references to talks at this CD12 Workshop will be presented in brackets}. Due to the fundamental nature of the pion and its interactions I decided that it would be timely in this outlook talk to compare where we are in $\pi\pi$ and $\pi$N scattering. This is my assessment of the status of this aspect of our field, and not a summary of the CD12 workshop.

Historically the pion was first postulated by Yukawa in 1938 as the quantum of the nucleon-nucleon interaction, and at the present time it is understood to dominate it at long range. Since the pion arises from chiral symmetry breaking it is a pseudoscalar and this leads e.g. to a non-central (tensor) interaction in the nucleon-nucleon interaction. In the context of QCD the pion plays a special role as the "signature" of chiral symmetry breaking. Therefore the study of its properties, decay modes, and its interaction with external fields and with other hadrons provides a crucial testing ground for low energy QCD. These observables can be calculated in ChPT and in lattice QCD. It is generally expected that QCD is correct and that any problems in testing it at the confinement scale are due to the difficulties in performing accurate calculations for this highly non-linear theory. This expectation should be tested experimentally which means we need to have reasonable estimates of the theoretical as well as the experimental errors and to reduce them. Both Lattice and ChPT calculations are proceeding in this direction as well as many dedicated experiments. In my view we are entering an era in which we can realistically discuss experimental tests of QCD at the confinement scale; for which the $\pi\pi$ and $\pi$N s wave scattering lengths and $\pi^{0}$ lifetime are good examples.

\section{Pion Properties}
\begin{figure}
\includegraphics[height=0.38\textwidth,width=0.5\textwidth] {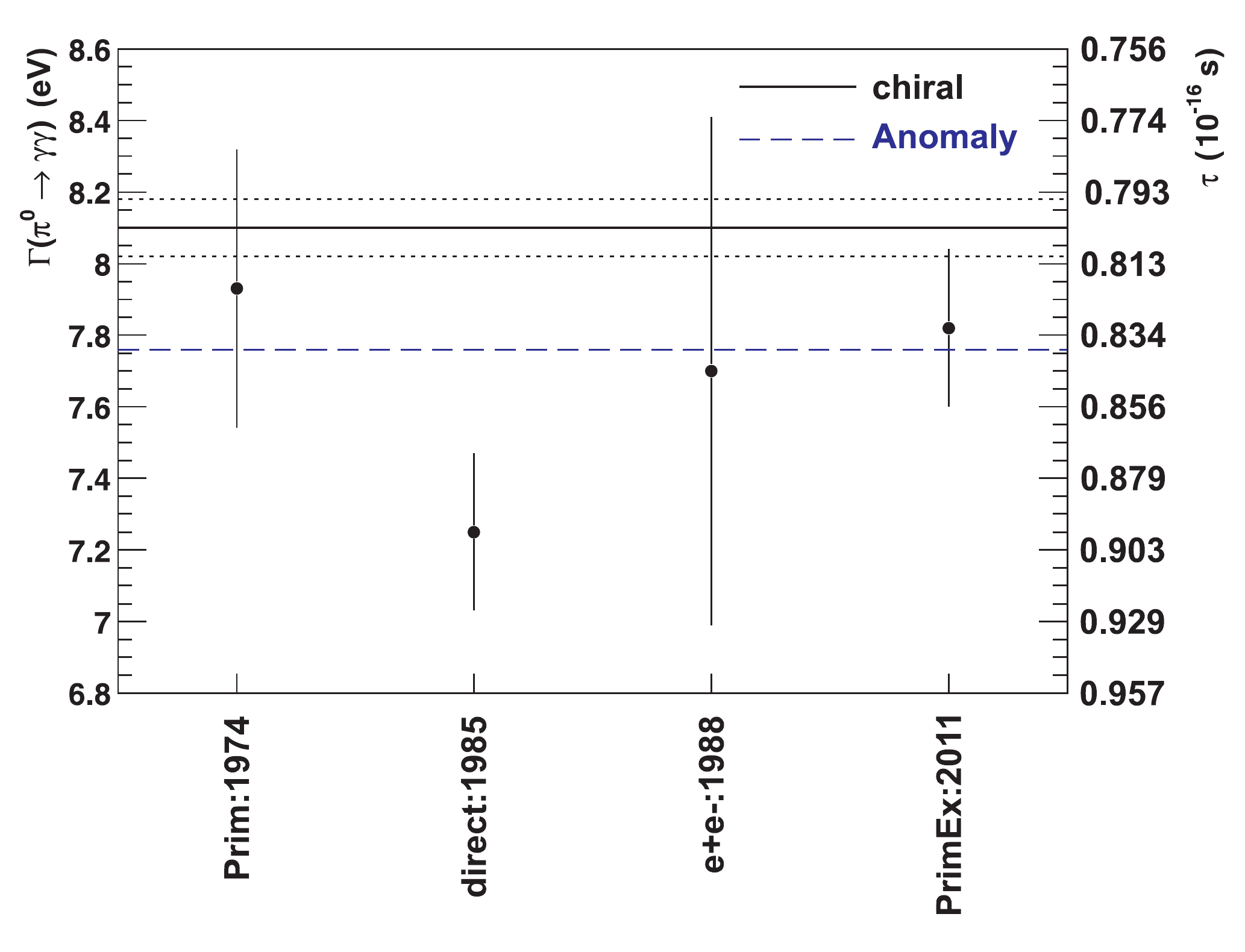}
\includegraphics[height=0.4\textwidth,width=0.48\textwidth] {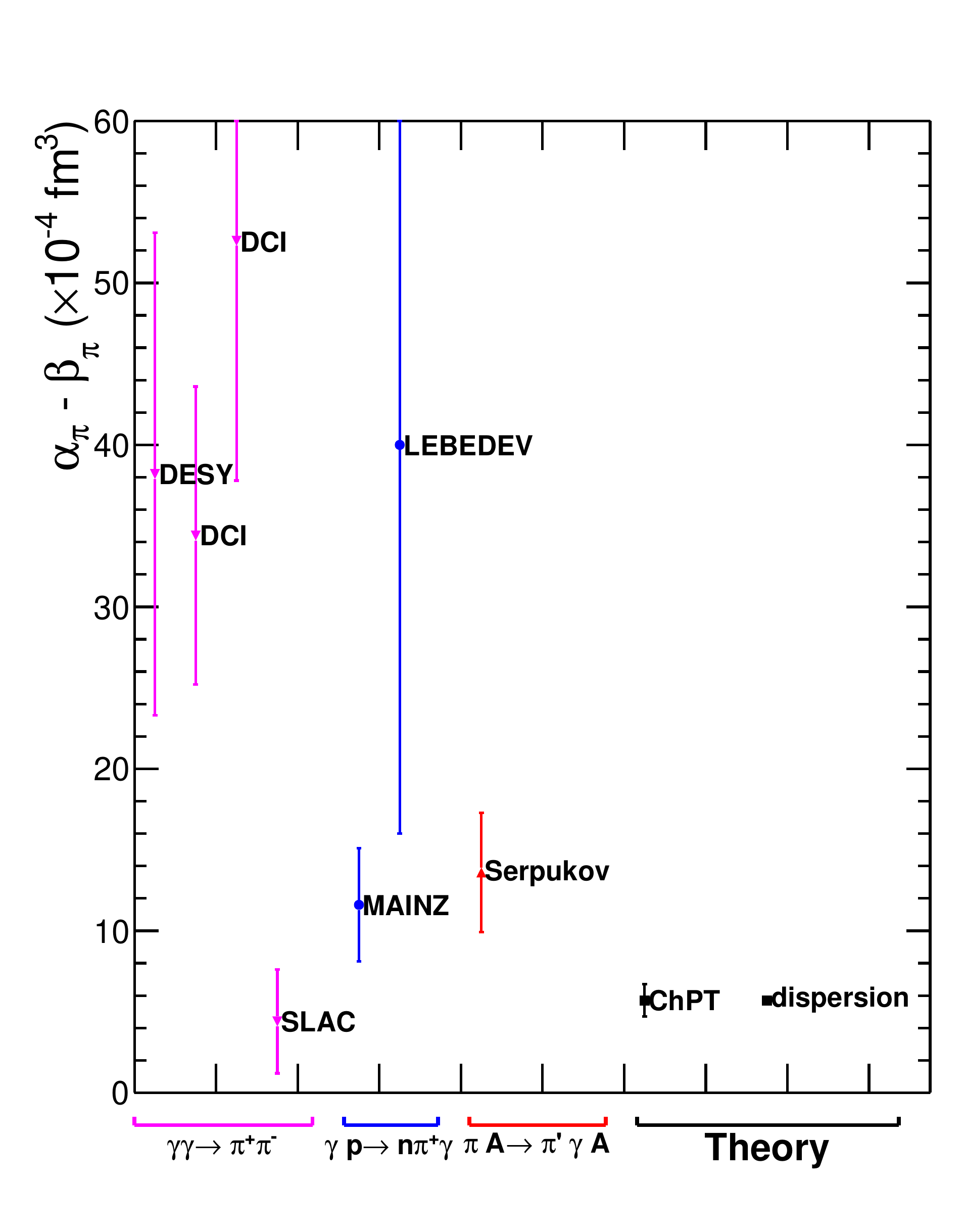}
\vspace{-0. cm}
\caption{Left Panel: The $\pi^{0}$ radiative decay width $\Gamma(\pi^{0} \rightarrow \gamma\gamma)$ in eV (left scale) and the $\pi^{0}$ mean life in units of $10^{-16}$ s (right scale) for the most accurate experiments (see\cite{RMP} for the references and details). The horizontal axis gives the technique and year of publication. The dashed horizontal line shows the prediction of the chiral anomaly\cite{anomaly} and the higher solid line shows the QCD (chiral) prediction with the 1\% error band\cite{Goity}. Right Panel: The polarizabilities of the charged pion, $\alpha_{\pi} -\beta_{\pi}$, in units of $10^{-4}$ fm$^{3}$. The first four measurements were performed with the $e^{+}e^{-} \rightarrow \gamma \gamma e^{+}e^{-} \rightarrow \pi^{+}\pi^{-} e^{+}e^{-}$ reaction; the next two with the $\gamma p \rightarrow \pi^{+}n\gamma$ reaction; and the next with the $\pi^{-} A \rightarrow \pi^{'-}\gamma A$ Primakoff reaction on a nuclear target. The two theoretical predictions use ChPT\cite{pi-pol-ChPT} and dispersion relations\cite{pi-pol-dispersion}. The data are summarized in\cite{pi-pol-ChPT}. Figure courtesy of  D. Lawrence.}
\label{fig:pion}
\end{figure}

  The $\pi^{0}$ meson is the lightest hadron since the electromagnetic interaction makes the $\pi^{\pm}$ mesons 4.6 MeV heavier. Its main decay mode is $\pi^{0} \rightarrow \gamma \gamma$ which is dominated by another QCD symmetry breaking effect, the axial anomaly\cite{anomaly}. The definition of an anomaly is when a symmetry of the classical Lagrangian is lost in the full quantum theory. For $\pi^{0}$ decay the conservation of the third isospin component of the axial current is lost upon quantization due to the fluctuations of the gauge fields. The lifetime predicted by the anomaly is exact in the chiral limit (when the light quarks are massless) and has no free parameters. This is the leading order term of the chiral series. At higher order (HO) the $\eta$ and $\eta^{'}$ mesons are mixed into the $\pi^{0}$ wave function. This is an isospin breaking effect and so is proportional to the light quark mass difference $m_{d} - m_{u}$\cite{Leut-mud}. This results in a $4.5 \pm 1$\% increase to the $\pi^{0}$ decay rate as calculated in chiral perturbation theory\cite{Goity}. Considering the fundamental nature of the subject, and the 1\% accuracy which has been reached in the theoretical lifetime prediction, it is important for experiments to aim for a comparable level of precision. A few years ago the particle data book average had an error of $\simeq$ 8\%, which perhaps was understated since many of the experiments were not in agreement at the quoted level of accuracy\cite{RMP}. In addition the experiments  were over 20 years old and could be improved with modern techniques. More recently an experiment using the Primakoff effect was performed at Jefferson Lab\cite{PrimEx}. The present status of the most accurate experimental results and theoretical predictions for the $\pi^{0}$ decay rate are shown in Fig.\ref{fig:pion}, based on our recent review of the $\pi^{0}$ lifetime\cite{RMP}. Three experimental methods have been used. One is the direct measurement of the decay distance of a high energy $\pi^{0}$. Two techniques measure the cross sections for $\gamma \gamma \rightarrow \pi^{0}$ production. In the Primakoff effect one of the gamma rays comes from the virtual Coulomb field of a target nucleus, and in colliding $e^{+}e^{-}$ beams both leptons produce virtual photons. As can be seen from Fig.\ref{fig:pion} the general magnitude of the prediction (the axial anomaly plus the increased rate due to the isospin breaking chiral corrections\cite{Goity}) has been confirmed. However the 1\% theoretical accuracy has not been matched. The two most accurate experiments are PrimEx with a quoted error of 2.8\% \cite{PrimEx} and the direct experiment at CERN with 3.1\% \cite{direct}. However, the difference between them is 7.5\%. This discrepancy clearly needs to be resolved if further progress is to be made. PrimEx has had a second run with plans to reduce the error by a factor of two. In addition plans are being considered by the Compass group to re-measure the "direct" experiment at the CERN SPS\cite{Future-pi0}. There are also plans to improve the $e^{+}e^{-}$ experiments at Frascati and BES\cite{Future-pi0}.
  
Important properties of the pion are its electric and magnetic polarizabilities. Although there are two accurate calculations of them based on ChPT\cite{pi-pol-ChPT} and dispersion relations\cite{pi-pol-dispersion} the measurements tend to be quite scattered as shown in Fig.\ref{fig:pion}. Space does not permit a discussion of these experiments. It is anticipated that the Compass collaboration is close to publishing a new, accurate, result based on the $\pi\gamma$ scattering using the Primakoff effect for 190 GeV $\pi^{-}$ mesons on Nickel targets. Preliminary results have been presented\cite{pi-pol-Compass} which are in agreement with the theoretical predictions\cite{pi-pol-ChPT,pi-pol-dispersion} with a total error of $\simeq 1\cdot 10^{-4}$ fm$^{3}$. This is based on a 2009 data run. A higher statistics run was taken in 2012 and the data are presently being analyzed\cite{pi-pol-Compass}. This is an excellent step in the determination of these fundamental quantities. It is also desirable to understand, if possible, why the previous experiments have obtained such widely differing results, since agreeing with the theory should not be a criterion for judging the validity of an experiment.

\section{$\pi\pi$ and $\pi$N Scattering}
Since the pion plays such a central role in chiral dynamics it is of interest to compare $\pi\pi$ and $\pi$N interactions, keeping in mind the restrictions imposed by the spontaneous hiding of chiral symmetry in QCD. To start we note that the $\pi$-hadron interaction is weak in the s wave near threshold and strong in the p wave at higher energies. This was first discussed in a pre-QCD seminal work by Weinberg where, using current algebra and PCAC, he calculated the s wave scattering lengths of pions interacting with any hadron\cite{W}. When applied to $\pi\pi$ and $\pi$N scattering the predictions are:
\begin{eqnarray}
a_{\pi\pi}^{I=0}= (7/4)L = 0.16/m_{\pi}, ~ ~a_{\pi\pi}^{I=2}= -(1/2)L = -0.0042/m_{\pi},~ ~L = m_{\pi}^{2}/(8\pi F_{\pi}^{2})
 \nonumber \\
 a_{\pi N}^{-} = (a_{\pi N}^{I=1/2}-a_{\pi N}^{I=3/2})/3= L/(1+m_{\pi}/M_{p})= 0.0795/m_{\pi} \nonumber \\ 
 a_{\pi N}^{+}= (a_{\pi N}^{I=1/2}+2a_{\pi N}^{I=3/2})/3= 0
 \label{eq:a_piN}
\end{eqnarray}
where I is the isospin, and $a_{\pi N}^{\pm}$ are the isoscalar and isovector $\pi$N scattering lengths. This remarkable result shows that $\pi$-hadron scattering goes to zero in the chiral limit (where the light quark masses $m_{u}$ and $m_{d}$, and therefore $m_{\pi}$, are zero) in agreement with Goldstone's theorem. These scattering lengths, $\simeq$ 0.1 fm, are approximately one tenth that of a typical hadron-hadron interaction of $\simeq$ 1 fm. Eq.\ref{eq:a_piN} does not have any dependence on the structure of the colliding pion and target, only on their mass and isospin. 

\begin{figure}
\includegraphics[height=0.35\textwidth,width=0.5\textwidth] {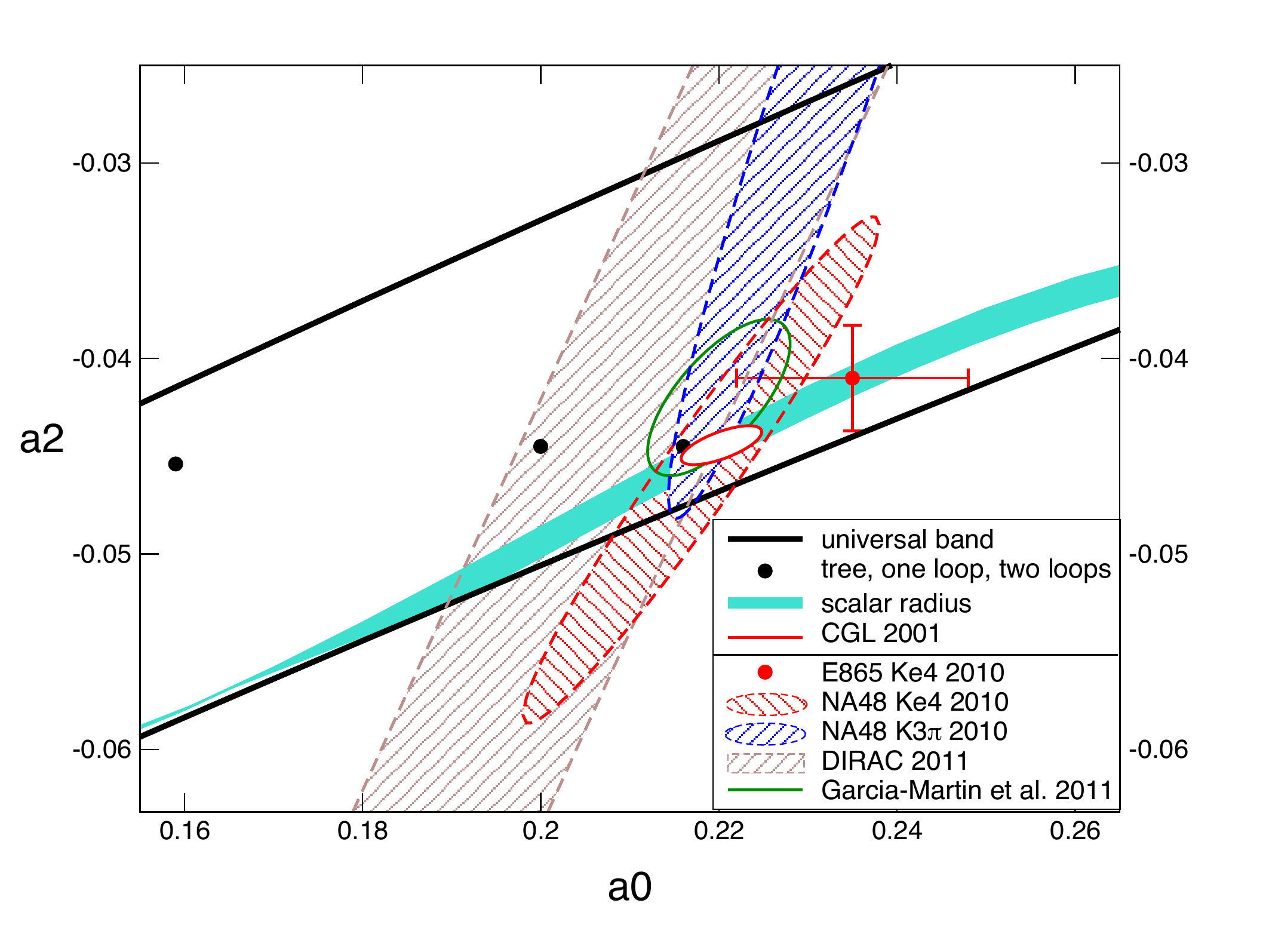}
\includegraphics[height=0.35\textwidth,width=0.5\textwidth] {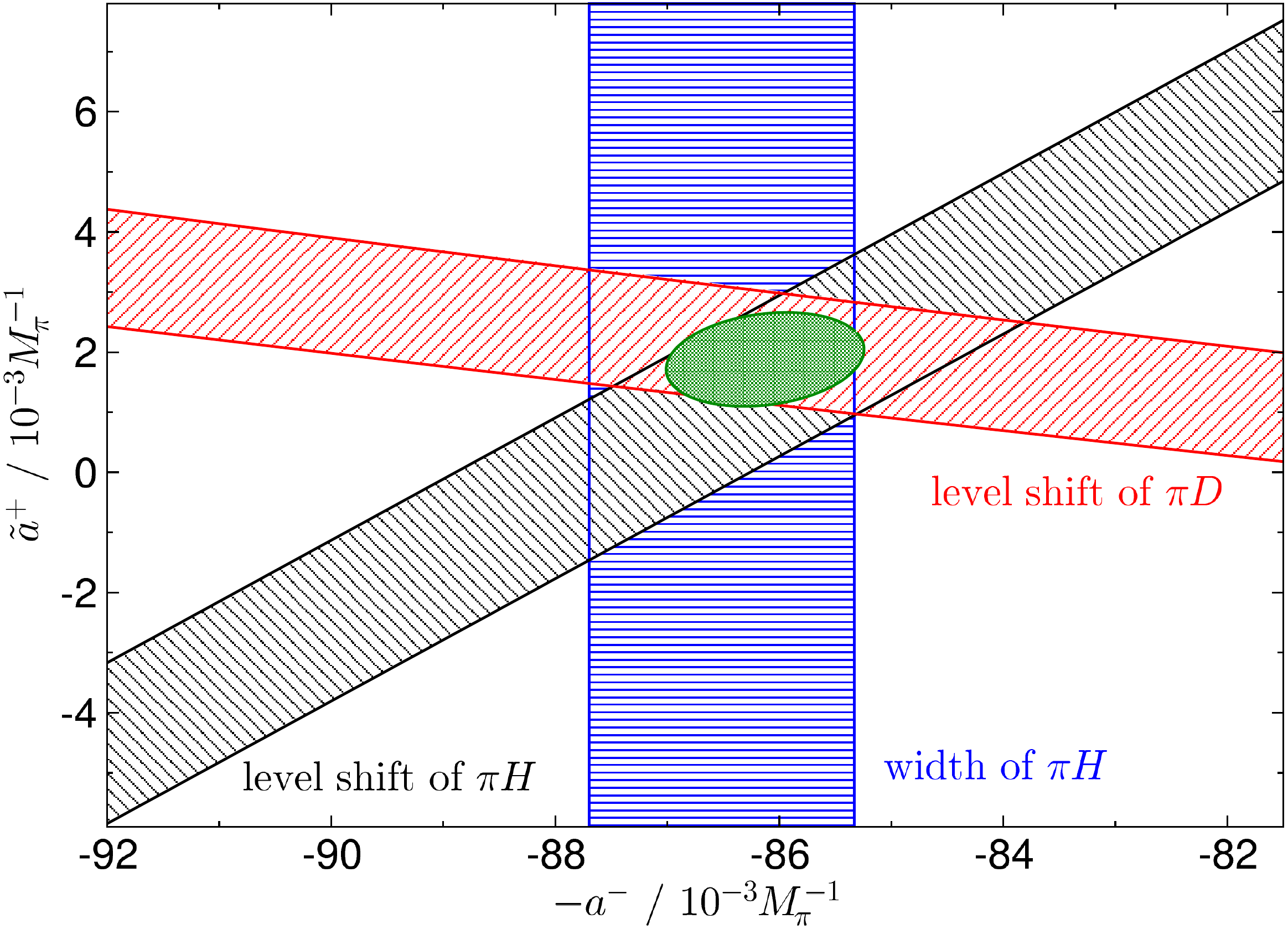}
\caption{Left Panel: Theoretical\cite{pi-pi} and experimental\cite{NA48}[Bizzeti] results for the I = 2 versus I = 0 $\pi \pi$ s wave scattering lengths in units of $1/m_{\pi}$. Figure courtesy of H. Leuwyler\cite{Leut-a_pi_pi}. Right Panel: $\pi$ N s wave scattering lengths in units of $10^{-3}/m_{\pi}$\cite{a_piN}. Figure courtesy of B. Kubis. See text for discussion.}
\label{fig:a_pi}
\end{figure}

It is now understood that Weinberg's calculation\cite{W} is the leading order $O(p^2)$ term in ChPT, an effective theory of QCD with an expansion in the momentum and pion mass\cite{ChPT,book,ChPT-review}. In the case of the $\pi\pi$ interaction ChPT calculations have been carried out to the next two orders\cite{pi-pi}. At the next to leading order, $O(p^4)$, $a_{\pi\pi}^{I=0}$ increased by 25\% from 0.16 to 0.20$/m_{\pi}$ due to the relatively rapid increase of the interaction strength with energy (technically through a chiral log). At the next order, $O(p^6)$, the increase of $a_{\pi\pi}^{I=0}$ is only to 0.216$/m_{\pi}$ indicating the convergence of the ChPT series. The accuracy of the ChPT calculations have been improved by incorporating dispersion relations\cite{pi-pi}. Ten years after they were made the accuracy of the experiments caught up. There have been a series of beautiful measurements by the NA48 collaboration at CERN\cite{NA48}, and were reported on in this workshop[Bizzeti]. They measured the final state $\pi\pi$ interactions in the $K_{e4}$ decays $K^{\pm} \rightarrow \pi^{+}\pi^{-} e^{\pm} \nu, K^{\pm} \rightarrow \pi^{0}\pi^{0} e^{\pm} \nu$. The results\cite{NA48}[Bizzeti] are in good agreement with the theoretical calculations\cite{pi-pi} as shown in Fig.\ref{fig:a_pi}.

There has also been a very impressive experimental effort at PSI to measure the level shifts and widths of pionic hydrogen and deuterium\cite{Gotta}, as well as a parallel theoretical effort\cite{a_piN}[Meissner]. The experimental accuracy has improved to the point where the isospin breaking effects due to the mass difference of the up and down quarks\cite{Leut-mud}, first pointed out by Weinberg\cite{W1}, had to be taken into account\cite{a_piN}[Meissner]; these were required to bring the $\pi$-N scattering lengths extracted from the pionic-hydrogen and deuterium results into agreement with each other. The final results, including the isospin breaking effects, are shown in Fig.\ref{fig:a_pi}. The results are $ a_{\pi N}^{-} = 0.0861 \pm 0.0009/m_{\pi}$ and $ a_{\pi N}^{+} = 0.0076 \pm 0.0031/m_{\pi}$. These are surprisingly close to the 1966 Weinberg calculation\cite{W}, particularly considering the number of higher order corrections.

\begin{figure}
\includegraphics[height=0.4\textwidth,width=0.38\textwidth] {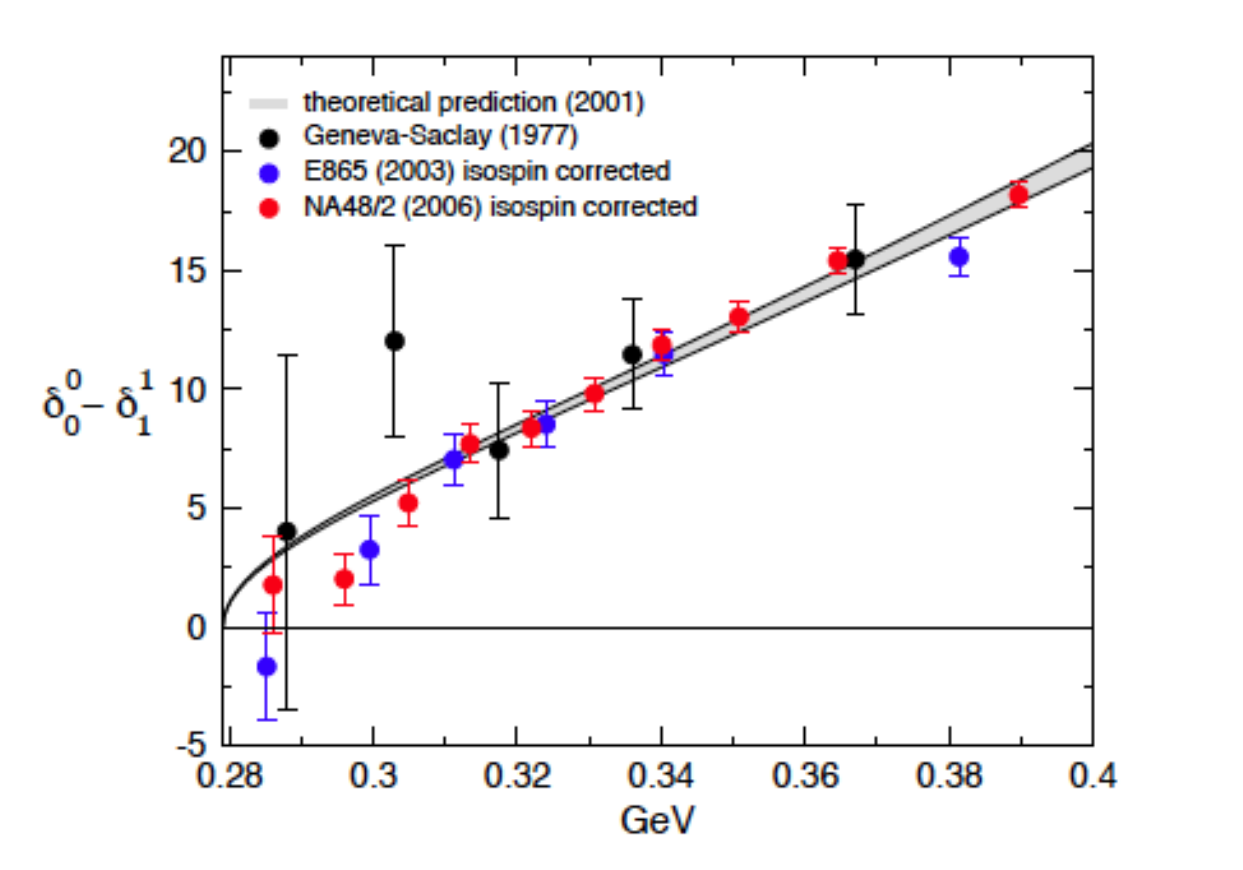}
\includegraphics[height=0.4\textwidth,width=0.27\textwidth] {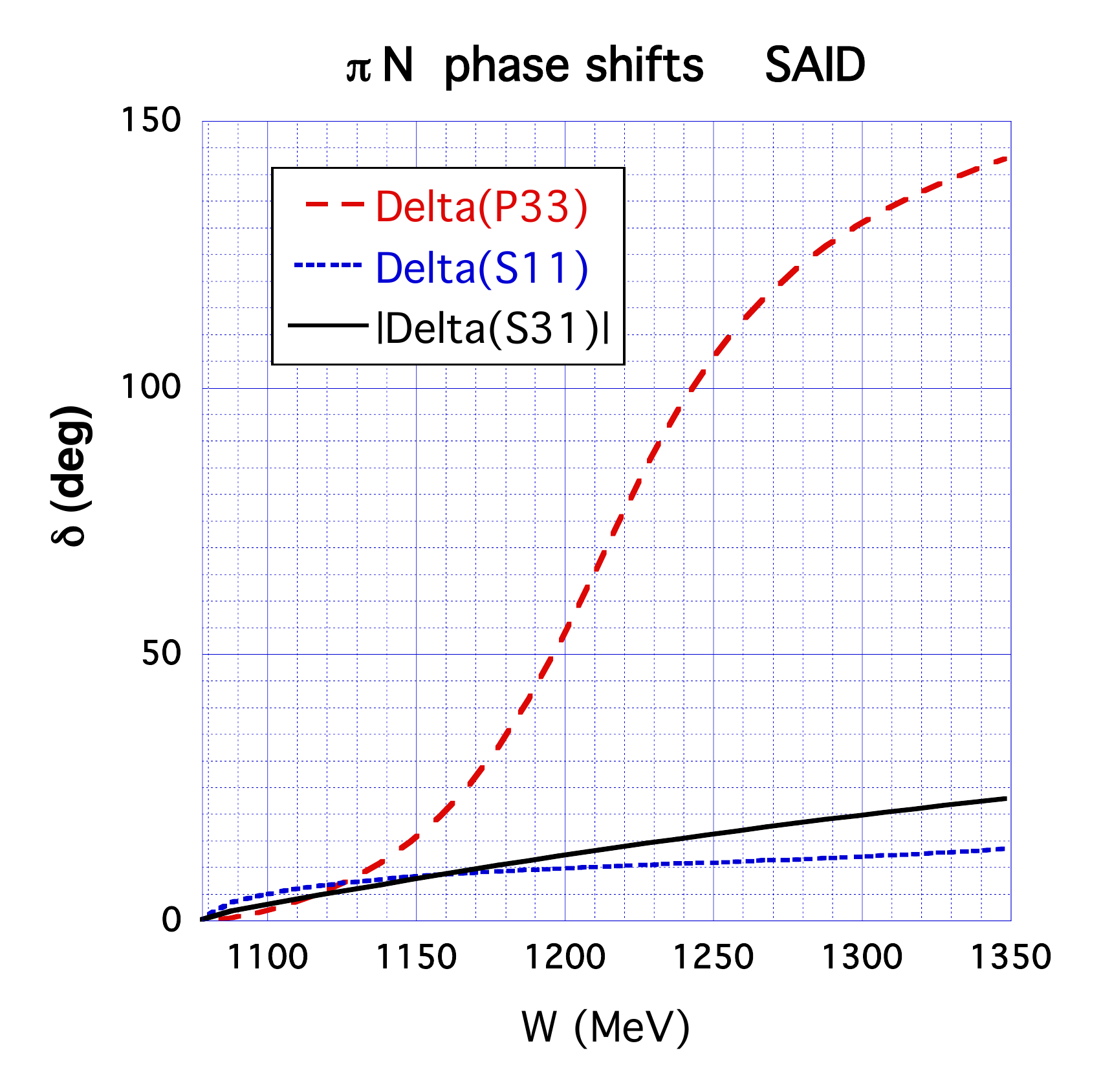}
\includegraphics[height=0.4\textwidth,width=0.34\textwidth] {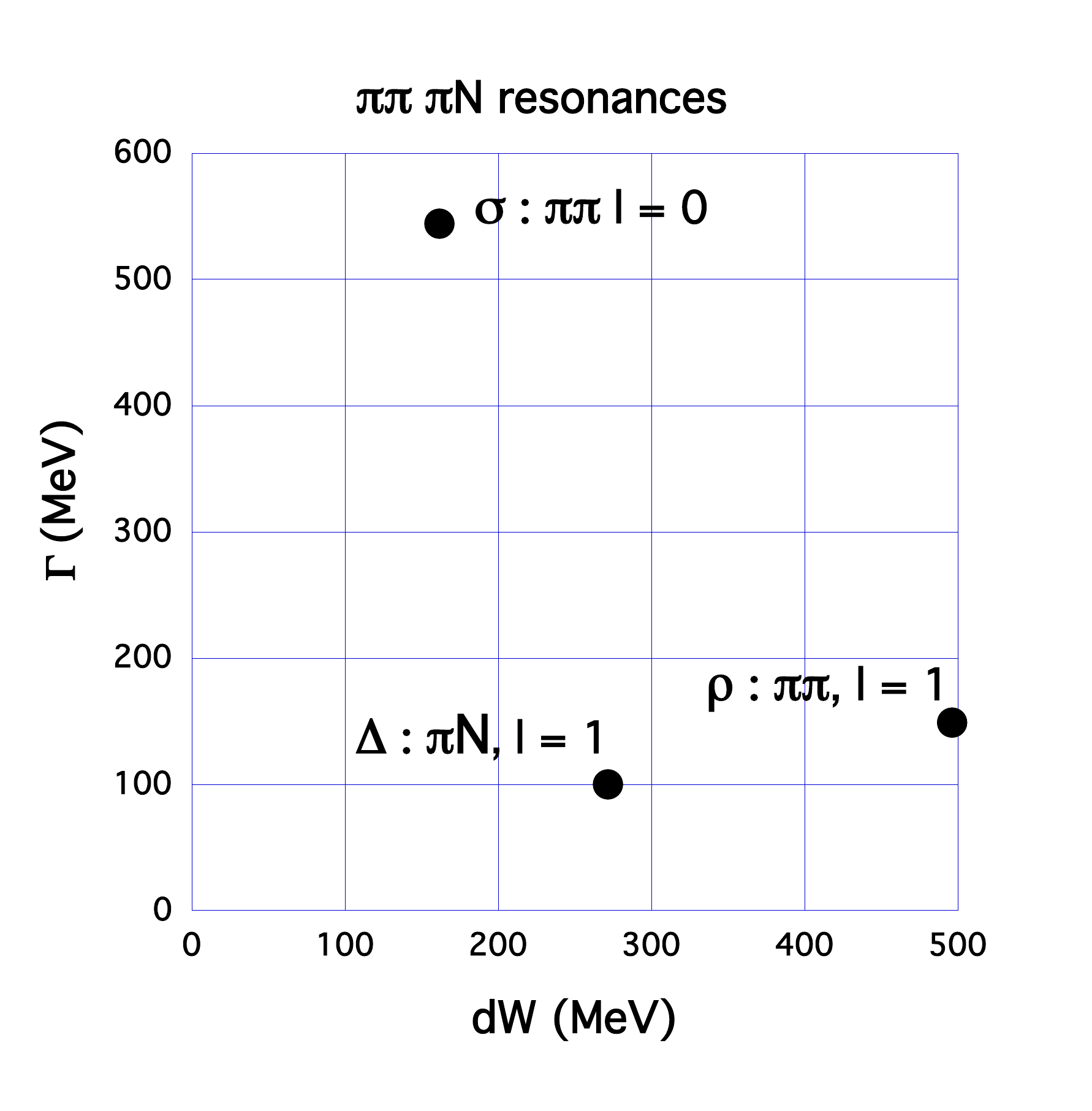}
\caption{Left Panel: I = 0 s wave phase shift $\delta_{0}^{0}$ minus p wave phase shift $\delta_{1}^{1}$ versus W in GeV for $\pi\pi$ scattering\cite{Leut-a_pi_pi}. Middle Panel: s and p wave phase shifts for $\pi$N scattering\cite{SAID}. Right Panel: Widths and energies of the poles in $\pi\pi$($\pi$N) scattering. $dW = W_{res} - 2m_{\pi}(W_{res}-M_{N})$ respectively.}
\label{fig:sp_phases}
\end{figure}

Going beyond the s wave scattering lengths to higher energies, the difference of the I = 0 s wave and I = 1 p wave phase shifts for $\pi\pi$ scattering, $\delta_{0}^{0}-\delta_{0}^{1}$, is presented in Fig.\ref{fig:sp_phases}. At low energies $\delta_{0}^{0} > \delta_{0}^{1}$ showing a relatively rapid rise in the I = 0 s wave $\pi\pi$ interaction which leads to the very broad $\sigma$ resonance at a mass of $441_{-8}^{+16}$, $\Gamma = 544_{-25}^{+18}$ MeV\cite{sigma} which  was explained qualitatively in terms of chiral dynamics\cite{pi-pi}. Fig.\ref{fig:sp_phases} shows the lowest lying resonances in the $\pi\pi$ and $\pi$N systems. The unusual $\sigma$ resonance is the closest to the ground state and has the largest width. The phase shift does not go through $90^{\circ}$ in the low energy region, and is larger than $\delta_{1}^{1}$ almost until the $\rho$ resonance. In the p wave the $\pi\pi$ interaction behaves in a more conventional way. At higher energies than is shown in Fig.\ref{fig:sp_phases} $\delta_{1}^{1}$ goes through $90^{\circ}$ at the $\rho$ resonance, W = 776 MeV, and has a width of $\simeq$ 150 MeV similar to most resonances. 

\begin{figure}
\begin{center}
\includegraphics[height=0.45\textwidth,width=0.48\textwidth] {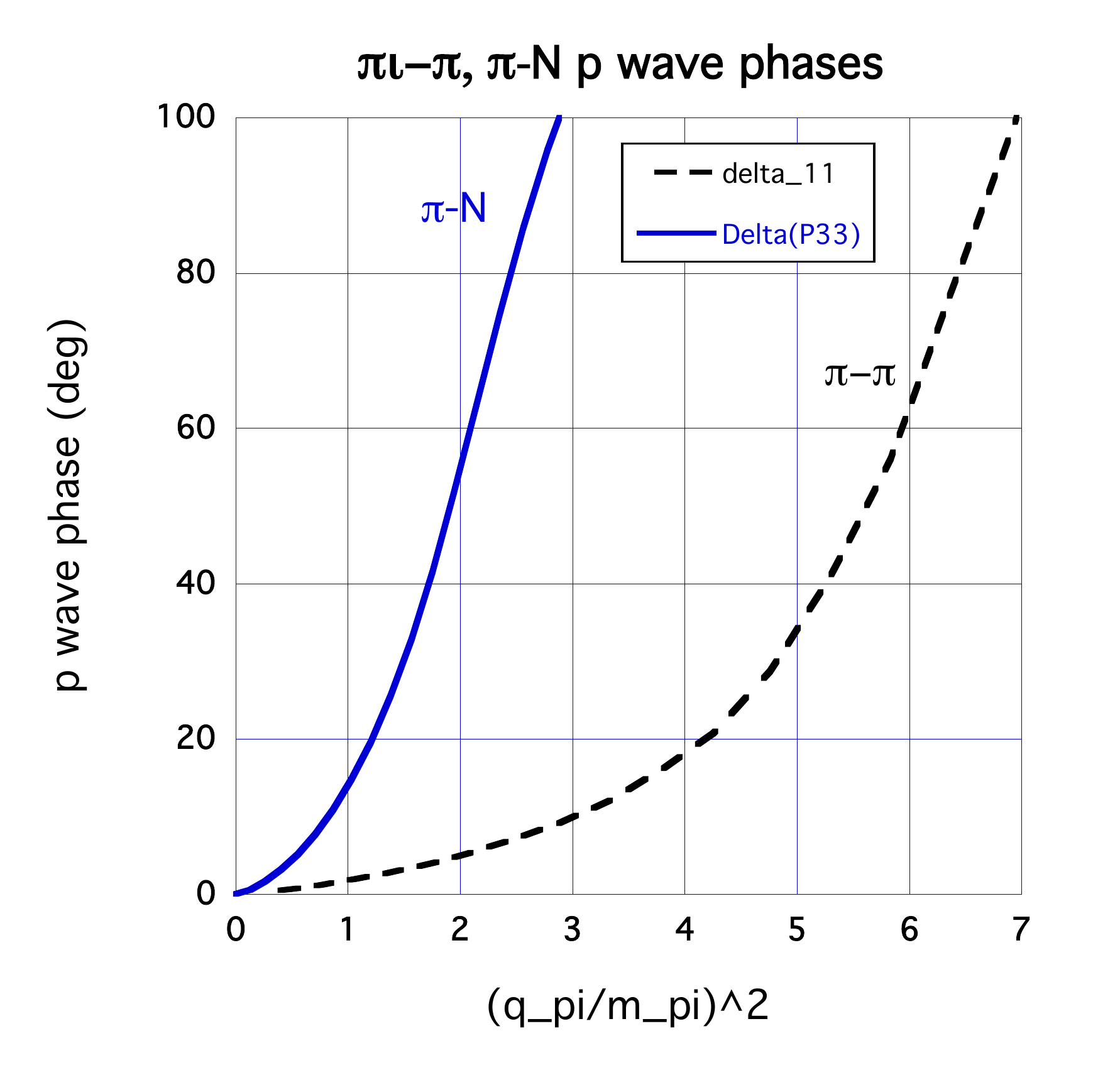}
\includegraphics[height=0.45\textwidth,width=0.48\textwidth] {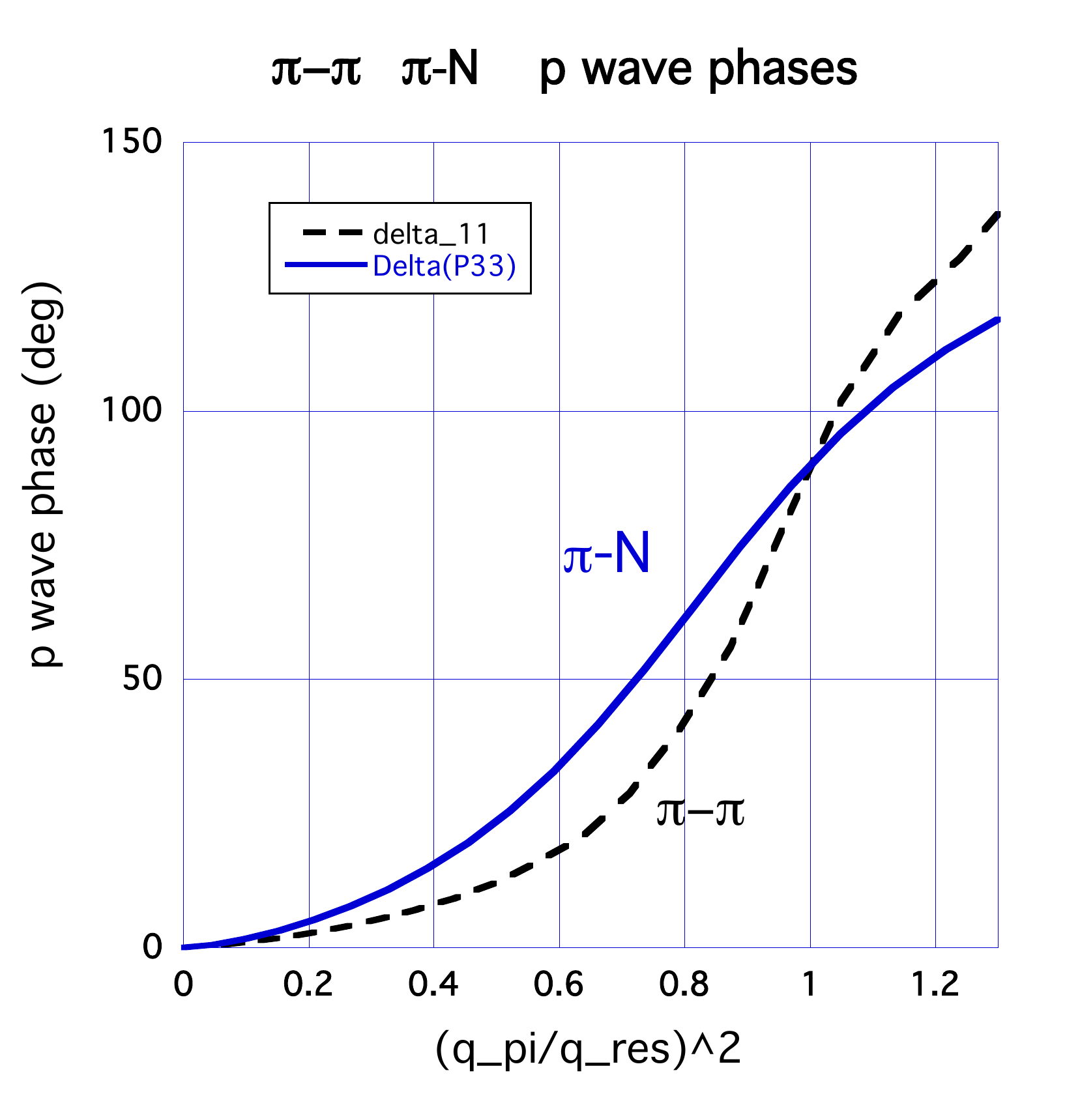}
\end{center}
\vspace{-1.0 cm}
\caption{Left Panel: Empirical p wave phase shifts versus the pion-CM momentum in units of $m_{\pi}$. The curves are for $\delta_{1}^{1}$ for $\pi\pi$ scattering\cite{pi-pi-empirical} and for $P_{33}$ for $\pi$N scattering\cite{SAID}. Right Panel: The same curves but with the pion-CM momentum relative to the decay CM momentum of the $\rho$ and $\Delta$ resonances. Numerical results for $\pi\pi$ scattering courtesy of J.R. Pelaez. 
 See text for discussion.}
\label{fig:p-phases}
\end{figure}

\begin{figure}
\begin{center}
\includegraphics[height=0.35\textwidth,width=0.45\textwidth] {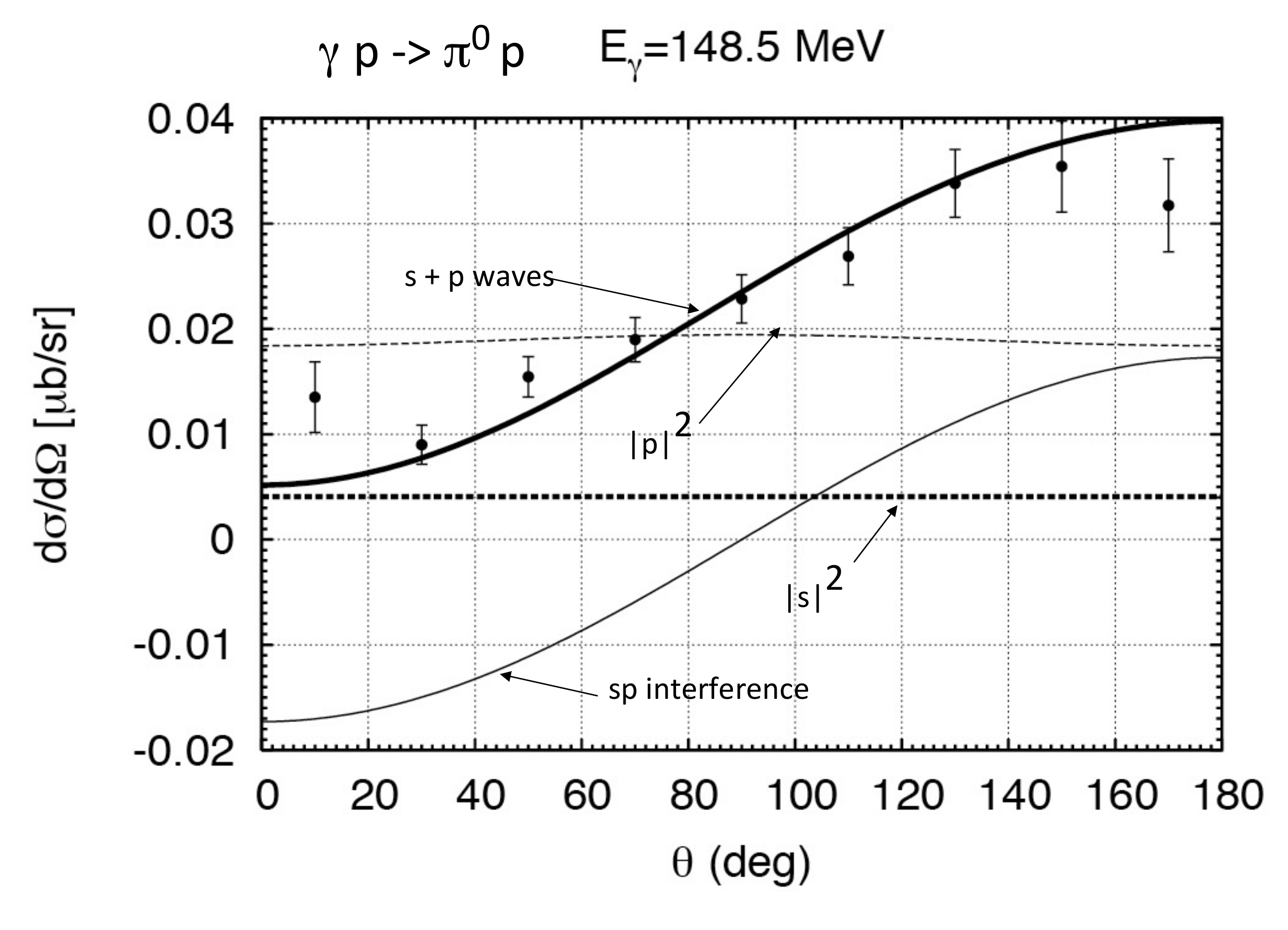}
\includegraphics[height=0.35\textwidth,width=0.45\textwidth] {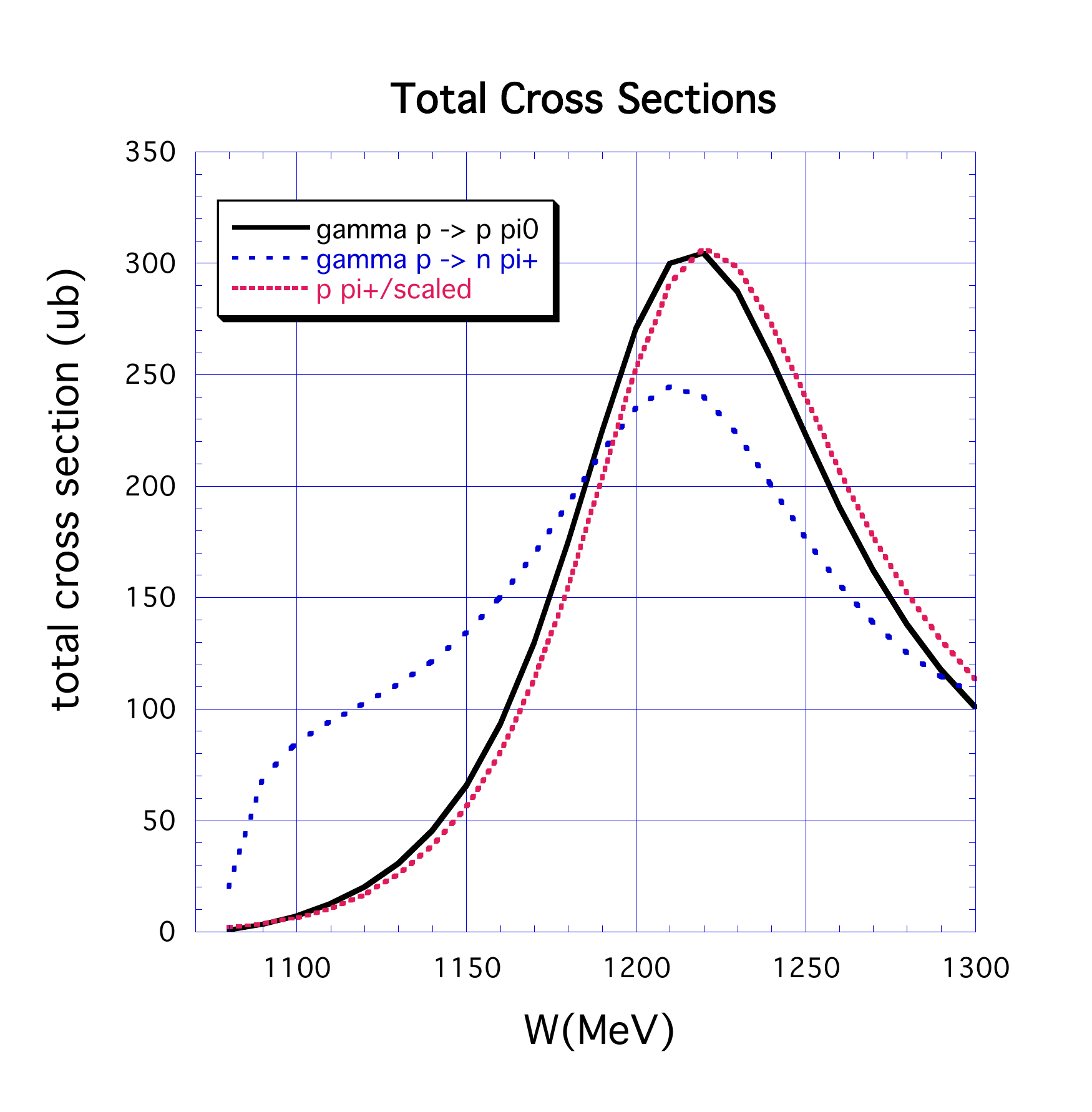}
\end{center}
\vspace{-1.0 cm}
\caption{Left Panel: The cross section for the $\gamma p \rightarrow \pi^{0}p$ reaction $\simeq$ 4 MeV above threshold showing the s and p wave contributions\cite{Hornidge,Cesar-fit}. Right Panel: Total cross section for the $\gamma p \rightarrow \pi^{0}p$, $\pi^{+}n$ reactions and $\pi^{+}p$ scattering (scaled to the same magnitude at the $\Delta$ resonance)\cite{Yangfest}. See text for discussion.}
\label{fig:sig_piN}
\end{figure}

Qualitatively the $\pi$N interaction behaves closer to what was expected, namely a small s wave scattering length as predicted by Weinberg\cite{W} and a strong p wave interaction leading to the $\Delta$ resonance pole at W = 1210 MeV, $\simeq$ 270 MeV above the nucleon mass. Fig.\ref{fig:sp_phases} shows that the $\Delta$ pole is closer to the ground state of the $\pi$N system than the $\rho$ is to the $\pi\pi$ ground state. The positions of these resonances, which cannot be calculated in ChPT, have a profound influence on the dynamics. The low lying $\Delta$ resonance means that the p wave amplitudes in both $\pi$N scattering and the $\gamma^{*} N \rightarrow \pi N$ reaction are more important, at low energies, than they are in $\pi\pi$ scattering. This is shown in Fig.\ref{fig:sp_phases}. In $\pi\pi$ scattering $\delta_{0}^{0} > \delta_{1}^{1}$, while for $\pi$N scattering the p wave phase shift is larger for the small values of dW $\geq$ 50 MeV. The relatively rapid rise of the p wave strength can be seen in the photo-pion data\cite{Hornidge,Cesar-fit}. As shown in Fig.\ref{fig:sig_piN}, by $\simeq$ 4 MeV above threshold the s wave amplitude is seen only in the sp interference with the already dominant p wave, producing a large forward-backward asymmetry in the differential cross section. A direct comparison of the p wave phase shifts for $\pi\pi$ and $\pi$N scattering is presented in Fig.\ref{fig:p-phases}. The phase is plotted versus the square of the pion-CM momentum since these are p waves and this is an intuitive way to take the centripetal barrier into account, at least for low energies. For $\pi$N scattering the $P_{33}$ phase shift (the resonant $\Delta$ channel, I = J = 3/2)\cite{SAID} rises rapidly in pion momentum compared to the $\delta_{1}^{1}$ phase for $\pi\pi$ scattering\cite{pi-pi-empirical}. This is in part due to the fact that $dW(\Delta) << dW(\rho)$. To try to approximately take the difference of dW into account the plot is also made in terms of the resonance decay CM momentum. To first approximation this works, though for momenta below the resonance the $\pi$N phase is still larger. This may be due to the fact that the chiral series for $\pi$N starts off linear in the momentum while the series for $\pi\pi$ scattering starts off as $p_{\pi}^{2}$. In addition large $N_{c}$ arguments indicate that the p wave interaction in $\pi$N  should be greater than in $\pi\pi$(see \cite{Nc} and references therein).

Another example of the relatively weak s wave at low energies and the strong p wave leading to the $\Delta$ resonance is in the total cross sections for $\pi^{+}p$ scattering and the $\gamma p \rightarrow \pi^{0}p$ reaction, demonstrated by Fig.\ref{fig:sig_piN}\cite{Yangfest}. These are often cited in textbooks as beautiful examples of a Breit-Wigner resonance shape. What is hardly ever mentioned is that this depends on the fact that the cross section is small at low energies. Fig.\ref{fig:sig_piN} also shows a dramatic example of how this picture can change when the cross section is large near threshold, as is the case for the $\gamma p \rightarrow \pi^{+}n$ reaction. Having charge in the final state dramatically increases the s wave magnitude (the Kroll-Ruderman term) and qualitatively changes the shape of the cross section\cite{Yangfest}.

During the past decade tremendous progress has been made in lattice calculations. They are being performed with pion masses close to the experimental values and lattice practitioners are doing their best to quantify their errors caused by lattice spacing, finite volume effects, and algorithms. For those of us on the periphery of this activity the work of the FLAG (Flavinet Lattice Averaging Group) collaboration is of special significance\cite{lattice-ChPT}[Lellouch], particularly in making connections between ChPT and lattice calculations. As the numerical solutions to QCD increase in accuracy they are able to test ChPT in some special ways. One is the off shell nature of the calculations such as the dependence of the pion mass on the light quark masses, which is not accessible to experiment. Another lattice test of ChPT is the determination  of the low energy constants (LECs) of ChPT. In the case of $\pi\pi$ scattering there are two LECs, $\bar{l_{3}}$ and $\bar{l_{4}}$; the first was determined in ChPT from $m_{\pi}$ and the second from the scalar pion radius or $F_{\pi}$\cite{pi-pi}. Lattice calculations determine these LECs from the light quark mass dependence of $m_{\pi}$ and $F_{\pi}$, and they are in agreement with the values determined by ChPT (see\cite{lattice-ChPT} Fig.9)[Lellouch]. This is consistent with ChPT calculations agreeing with QCD! 

\begin{figure}
\begin{center}
\includegraphics[height=0.35\textwidth,width=0.5\textwidth] {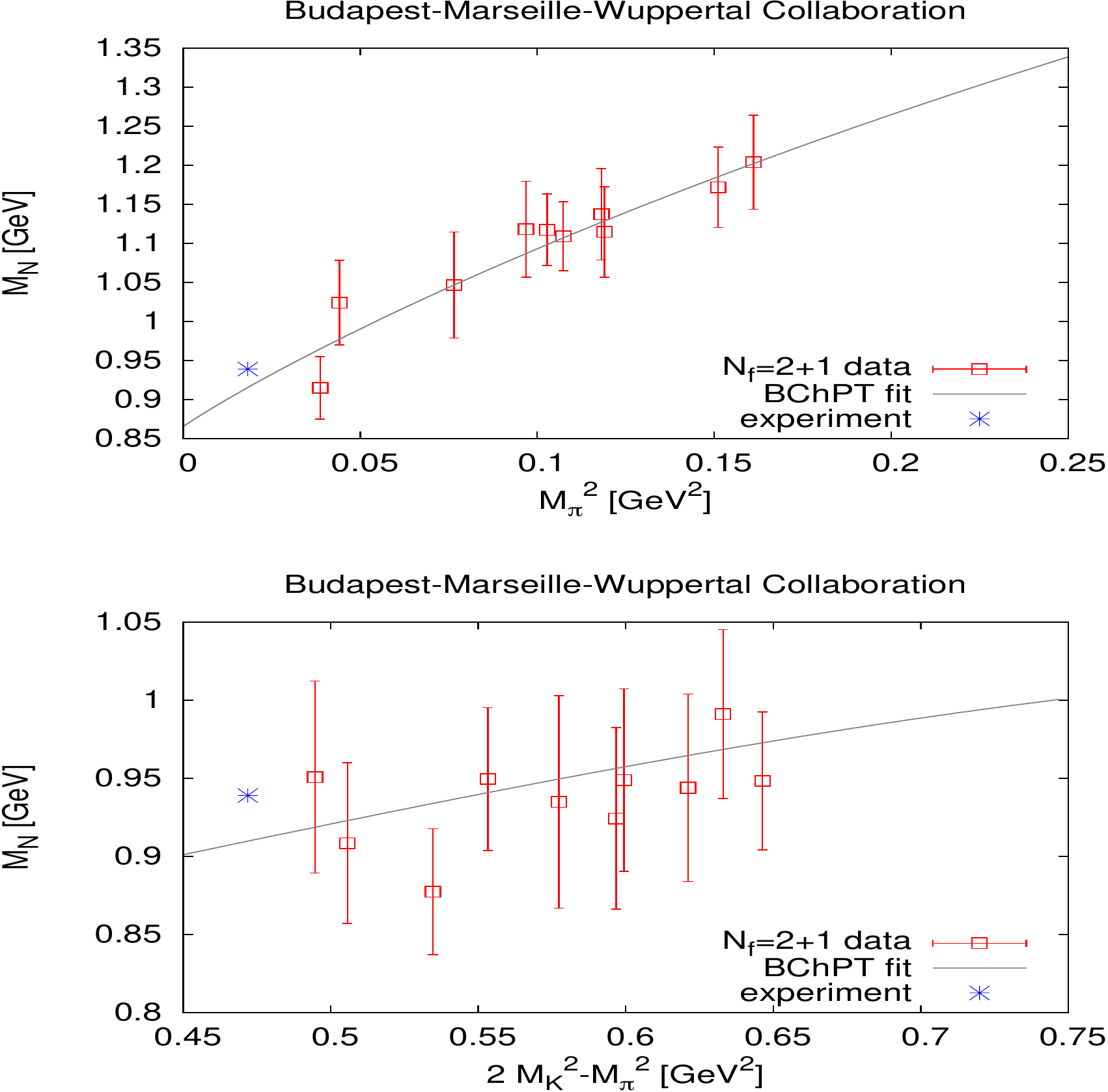}
\includegraphics[height=0.35\textwidth,width=0.45\textwidth] {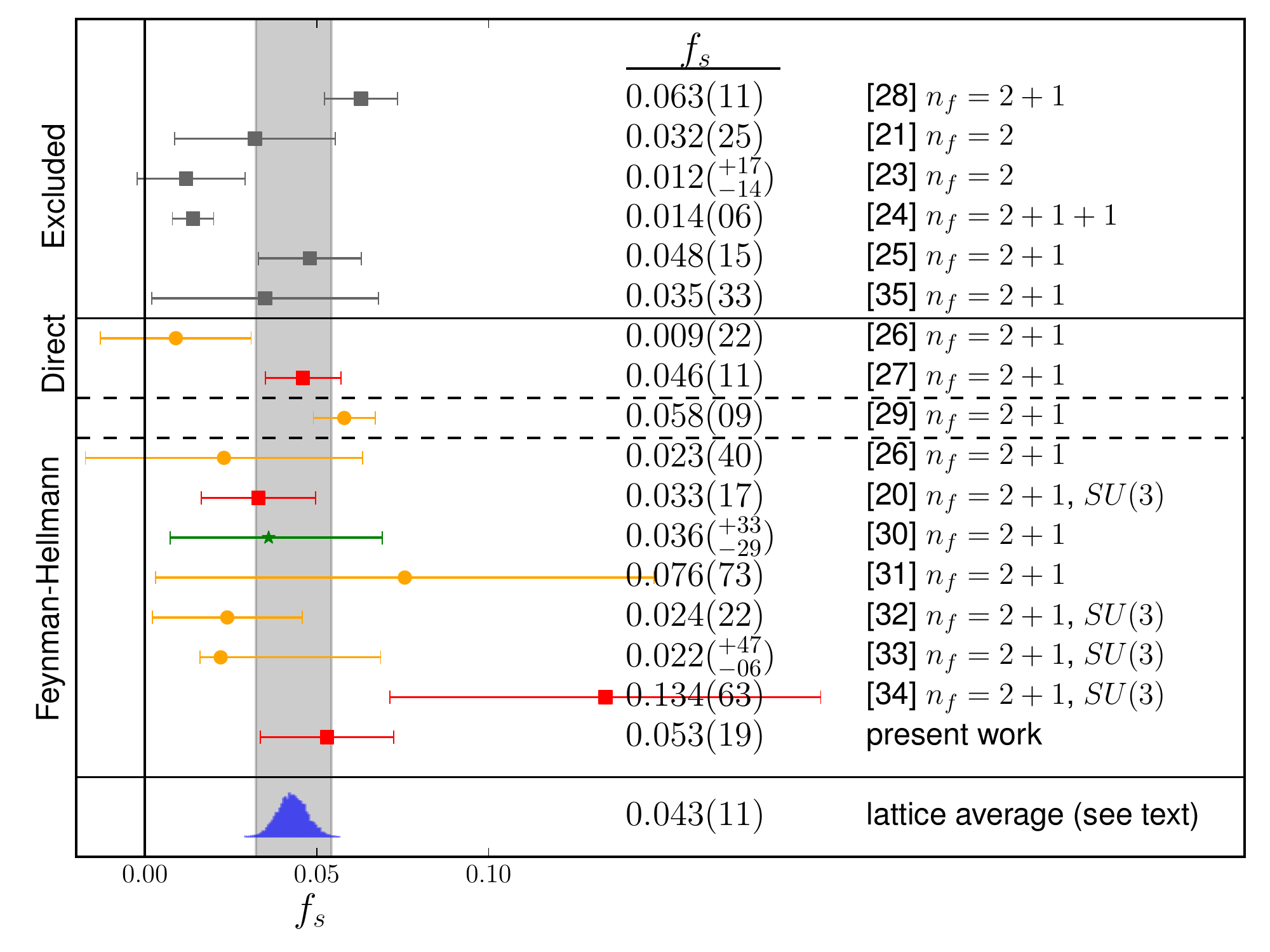}
\end{center}
\vspace{-0.5 cm}
\caption{Lattice Calculations: Left Top Panel: Nucleon mass $M_{N}$ versus $m_{\pi}^{2}$. Left Bottom Panel: $M_{N}$ versus 2$m_{K}^{2}-m_{\pi}^{2}$. Right Panel: Averaging for $f_{S}$. Figures courtesy of S. Durr and A. Walker-Loud. See text for discussion.}
\label{fig:f_s}
\end{figure}

In the nucleon sector there has been a long standing but inconclusive effort to determine the $\pi$N $\sigma$ term and the contribution of the strange quark to the mass of the proton\cite{pi-N-sigma}. These efforts have used ChPT fits to the data often including dispersion relations to extrapolate the results into the unphysical region (the Cheng-Dashen point) where theory meets experiment\cite{pi-N-sigma}. This subject was discussed at this workshop[Camalich, arXiv:1303.3854, Walker-Loud]. Recent lattice calculations have achieved reasonable accuracy for the $\pi$N $\sigma$ term, as illustrated in Fig.\ref{fig:f_s}. The calculations of the nucleon mass $M_{N}$ versus $m_{\pi}^{2}$ are used to extract the $\pi$N $\sigma$, term and $M_{N}$ versus $2 m_{K}^{2} - m_{\pi}^{2}$ are used to obtain the strange quark contribution to $M_{N}$\cite{Durr}. The fractional contribution, $f_{s}$, of the strange quark mass to the nucleon mass is shown in Fig.\ref{fig:f_s}. This shows the values obtained by lattice calculations where, despite the scatter of the individual points, an average value of $f_{s} = 0.043 \pm 0.011$ was obtained\cite{Andre-fS}. This appears to be reasonably accurate and  physically acceptable value. It is important to continue both the lattice effort and the ChPT approach with modern data to determine both the value of the $\pi$N  $\sigma$ term and whether the two methods are in agreement. 

I conclude the comparison of $\pi\pi$ and $\pi$N scattering by noting that indeed these interactions  have similarly small s wave scattering lengths as first predicted by Weinberg\cite{W}. They also both have strong p wave interactions leading to the $\rho$ and $\Delta$ resonances. However, starting right above threshold they have significant differences. Some of the differences are "imposed" by the positions of the resonant poles which are beyond the scope of ChPT, although closely related to chiral symmetry hiding. Some of the differences are due to the fact that the nucleon has spin 1/2 whereas the pion has spin 0. The Born terms in $\pi$N scattering (which are not present in $\pi\pi$ scattering) do not contribute to the $\pi$N s wave scattering lengths but do change the phases at higher energies. The fact that the nucleon and pion are not identical is important as different quantum states are involved. In addition there is the mass of the nucleon which is almost independent of the light quark masses. By contrast the pion mass goes to zero in the chiral limit. So for these reasons, and more, one might expect differences between $\pi\pi$ and $\pi$N scattering to even be larger than found empirically and sketched here. For me, making this comparison has been very interesting and I'm motivated to look into this more deeply. 

\section{Electromagnetic Pion Production from the Nucleon: A New Era} 

Electromagnetic pion production $\gamma^{*} N \rightarrow \pi N$ with real and virtual photons both compliments $\pi$N scattering and charge exchange reactions, and adds several important new features\cite{AB-review}. Measurements of the electromagnetic production amplitudes provide new tests for theoretical calculations, including strong isospin breaking due to the mass difference of the up and down quarks $m_{d}-m_{u}$, which is an important frontier of present studies. Although ChPT calculations have been carried out for isospin breaking in $\pi$N scattering\cite{W,a_piN,Martin} they have not been extended to the EM production amplitudes. For the $\gamma^{*} p \rightarrow \pi^{0}p$, $\pi^{+}n$ reactions charge states are reached which are different than with $\pi^{\pm}p$ reactions, enabling isospin breaking tests. Measurements which are sensitive to the final state $\pi$N interactions\cite{AB-lq} are possible with time reversal odd observables using transverse polarized targets in the $\gamma \vec{p} \rightarrow \pi N$ reactions and with the $TL^{'}$ structure function in the $\vec{e} p \rightarrow e^{'} \pi N$ reactions. Electromagnetic pion production is capable of starting at threshold and providing continuous energy coverage. For pion beams experiments are limited to pionic atoms initiated with stopped $\pi^{-}$ beams, or with $\pi^{\pm}$ kinetic energies above $\simeq$ 25 MeV due to the pions decaying in flight. 

\begin{figure}
\begin{center}
\includegraphics[height=0.4\textwidth,width=0.3\textwidth] {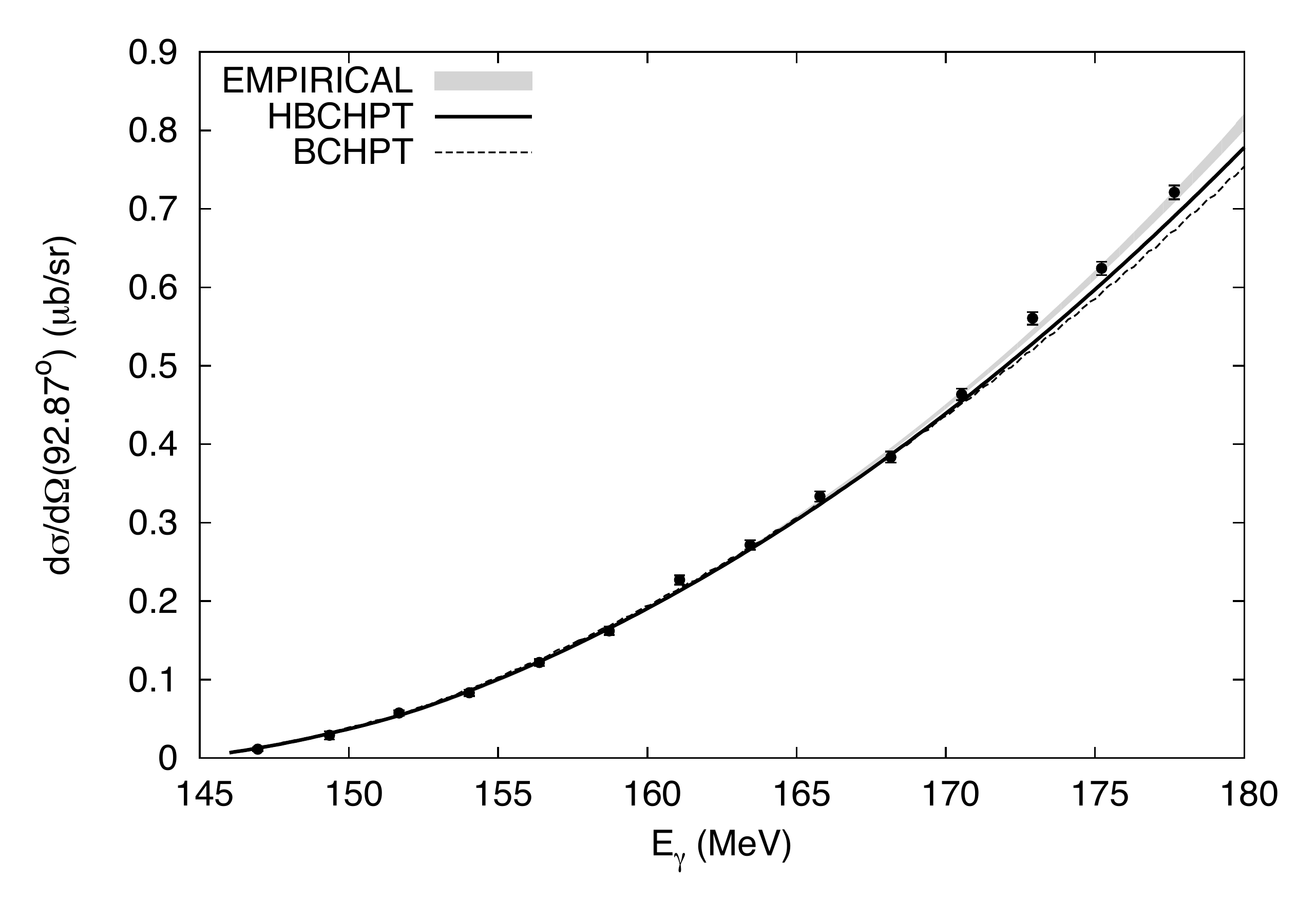}
\includegraphics[height=0.4\textwidth,width=0.3\textwidth] {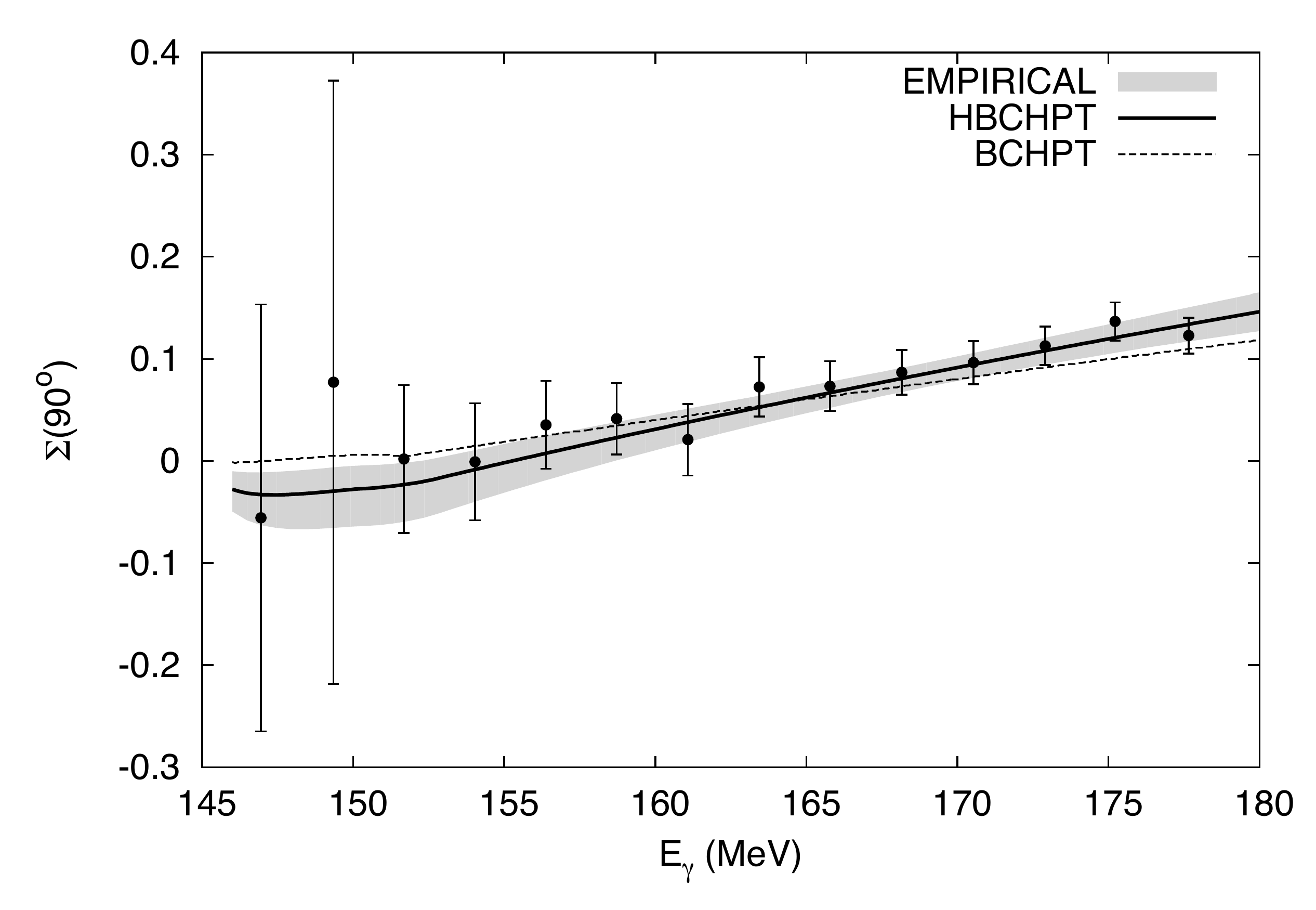}
\includegraphics[height=0.38\textwidth,width=0.35\textwidth]{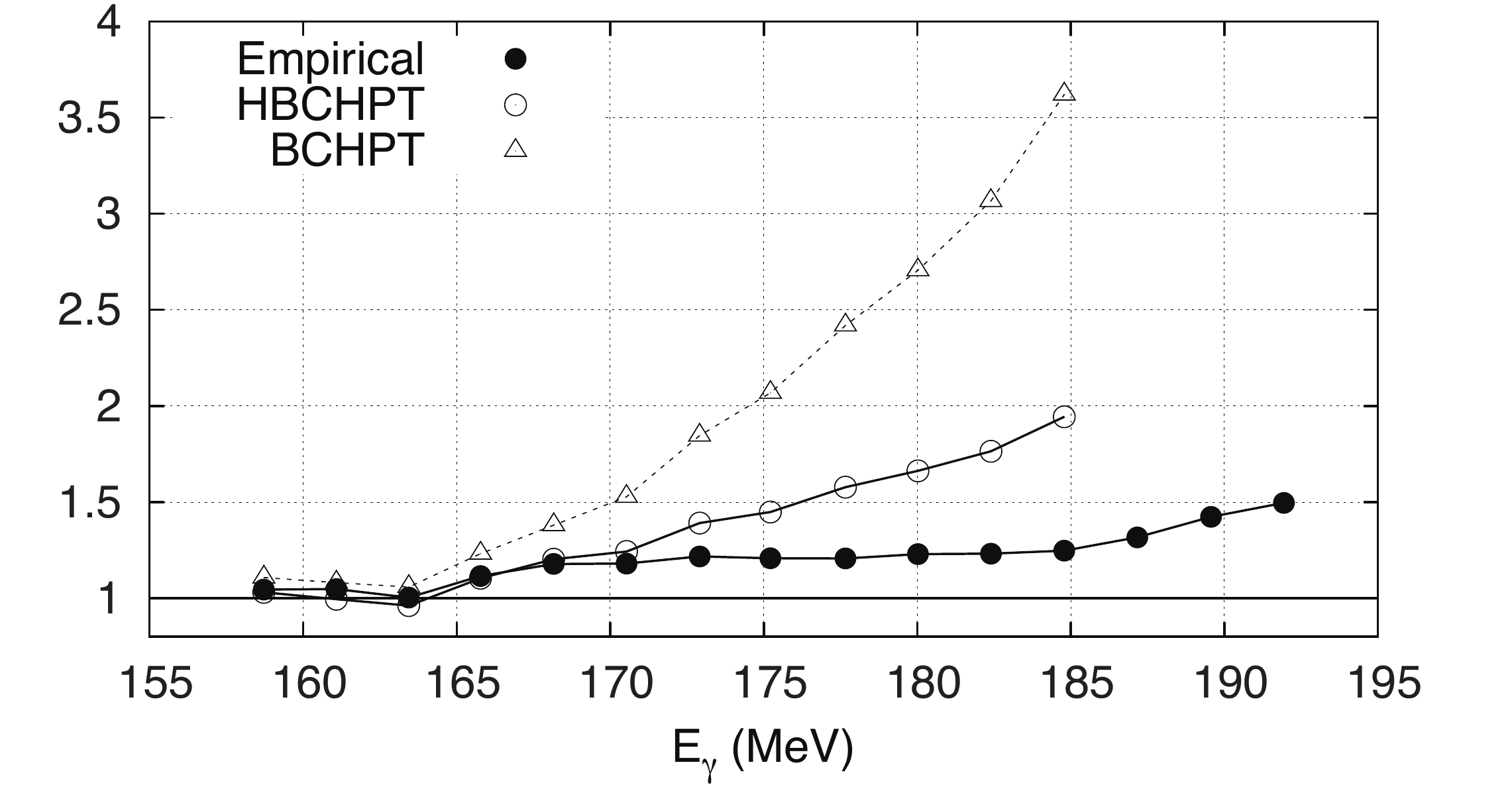}
\end{center}
\vspace{-0.8 cm}
\caption{Left Panel: Differential cross section at $93^{\circ}$ for the $\gamma p \rightarrow \pi^{0}p$ reaction as a function of photon energy\cite{Hornidge}. Middle Panel: The polarized photon asymmetry $\Sigma$ at $90^{\circ}$ versus photon energy for the $\gamma p \rightarrow \pi^{0}p$ reaction\cite{Hornidge}. The curves are the HBChPT\cite{Cesar-fit}, relativistic ChPT (BChPT)\cite{Hilt}, and empirical (with its error band)\cite{Cesar-fit} fits. Right Panel: $\chi^{2}$ per degree of freedom for the HBChPT, relativistic ChPT (BChPT) and empirical fits to the $\gamma p \rightarrow \pi^{0}p$ reaction as a function of the maximum energy of the fit\cite{Cesar-fit}.}
\label{fig:obs}
\end{figure}

\begin{figure}
\begin{center}
\includegraphics[height=0.35\textwidth,width=0.45\textwidth] {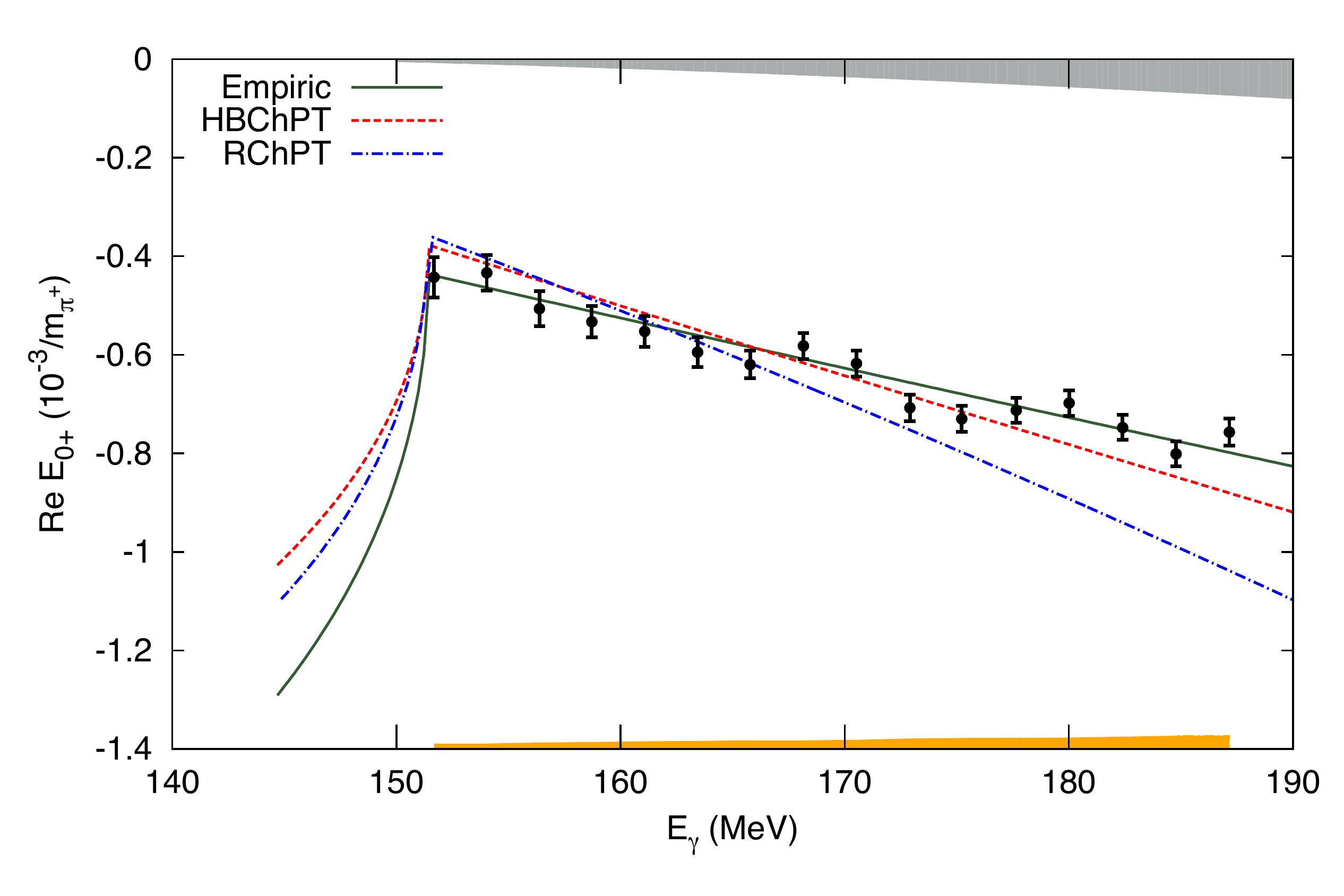}
\includegraphics[height=0.35\textwidth,width=0.45\textwidth] {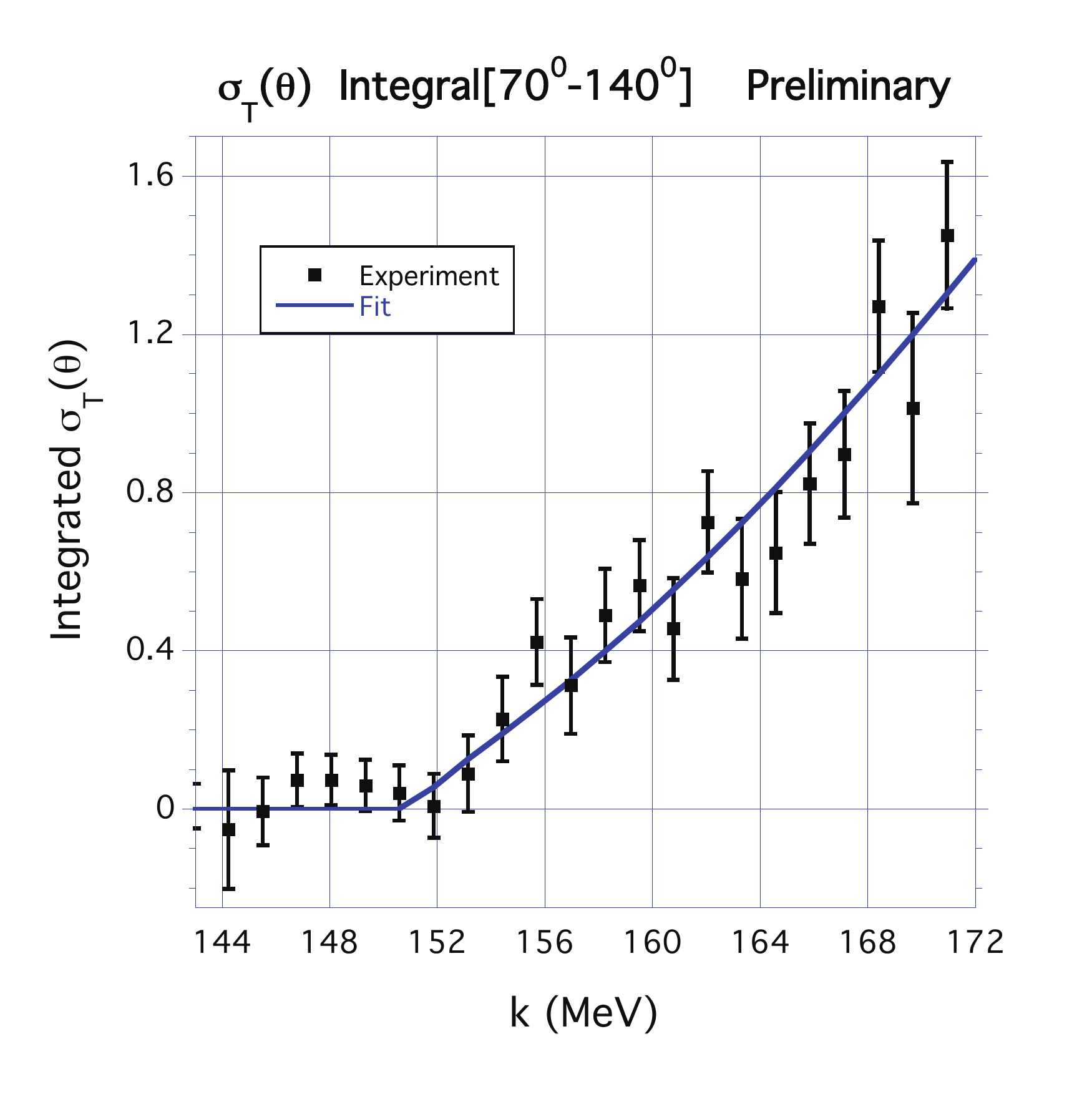}
\end{center}
\vspace{-1.0 cm}
\caption{Left Panel: Real part of the s wave multipole Re$E_{0+}$ for the $\gamma p \rightarrow \pi^{0}p$ reaction as a function of photon energy\cite{Hornidge}. Right Panel: Preliminary results for the integrated cross section for the transverse polarized target cross section $\sigma_{T}$ from $70^{\circ}-140^{\circ}$ (in arbitrary units) versus photon energy for the $\gamma p \rightarrow \pi^{0}p$ reaction\cite{Hornidge}. The curve is an empirical fit.}
\label{fig:E0p}
\end{figure}

The A2 (real photon) collaboration at Mainz is performing photo-pion experiments with tagged, polarized, photons\cite{Hornidge}[Hornidge] from threshold to above the $\Delta$ region with polarized targets\cite{A2-prop} and almost 4$\pi$ coverage. Some results of these experiments are shown in Fig.\ref{fig:obs}. We have achieved both small errors on the cross sections and the first measurement of the polarized photon asymmetry $\Sigma$ as a function of photon energy. The curves are calculations using Heavy Baryon Chiral Perturbation Theory (HBChPT)\cite{loop} with the low energy constants LECs updated with fits to this data\cite{Cesar-fit}[Fernandez-Ramirez], relativistic ChPT (labeled BChPT)\cite{Hilt}, and an empirical fit to the data\cite{Cesar-fit}. These calculations all include the contributions from the d waves calculated in the Born approximation\cite{Cesar-D}. The values of $\chi^{2}$/DOF as a function of the maximum energy to which the fit was performed is also presented in Fig.\ref{fig:obs}. The relativistic ChPT and HBChPT calculations agree with the data up to a maximum energy of $\simeq$ 165 to 170 MeV, $\simeq$ 25 MeV above threshold. This is the first systematic test in the $\gamma \pi$N sector of agreement between ChPT and experiment as a function of energy. The electromagnetic pion production amplitudes have also been extracted from the data as a function of energy\cite{Hornidge,Cesar-fit}. An example of this for the real part of the electric dipole amplitude Re$E_{0+}$ is presented in Fig.\ref{fig:E0p}. The agreement between the calculated multipoles and the empirically extracted ones are good up to the maximum energy of $\simeq$ 165 to 170 MeV, as was true for the observables.

An important ingredient of the success obtained with chiral calculations of $\pi\pi$ scattering is incorporating unitarity and also dispersion relations to take higher order terms into account\cite{pi-pi}. A similar development is underway in $\pi$N scattering calculations incorporating dispersion relations, unitarity and crossing symmetry[Dissche]\cite{Martin-dispersion}. Another important new calculation has been performed with a chiral Lagrangian to $O(q^{3})$ with dispersion relations to take higher order terms into account\cite{Lutz}; the coupling between the three active channels, Compton, neutral and charged pions, is taken into account by imposing unitarity.
The low energy constants were fit at higher energies, and reasonable agreement with pion scattering and photo-production is obtained over a wider energy region (W up to $\simeq$ 1300 MeV) than conventional ChPT\cite{Lutz}. The predictions, made before the present photo-pion experiment data were analyzed, are not in good  agreement with the new data\cite{Hornidge}. It remains to be seen if re-fitting the low energy constants will improve the agreement with this new data without spoiling the overall general agreement with the higher energy data.

The sharp downward break in Re$E_{0+}$ below the $\gamma p \rightarrow \pi^{+}n$ threshold $k_T(\pi^{+}n)$ at 151.4 MeV (Fig.\ref{fig:E0p}) is due to a unitary cusp\cite{AB-lq}. This is caused by the interference between the two quantum paths to the final state, namely the direct $\gamma p \rightarrow \pi^{0}p$ and the two-step $\gamma p \rightarrow \pi^{+}n \rightarrow \pi^{0}p$ amplitudes\cite{AB-lq}. An accurate measurement of this downward slope will enable an extraction of the cusp parameter $\beta = E_{0+}(\gamma p \rightarrow \pi^{+} n) a_{cex}(\pi^{+}n \rightarrow \pi^{0}p)$. Since Re$E_{0+}(\gamma p \rightarrow \pi^{+}n)$ has been measured independently in the cross section for the $\gamma p \rightarrow \pi^{+}n$ reaction, measuring $\beta$ will enable us to obtain $a_{cex}(\pi^{+}n \rightarrow \pi^{0}p)$, a measurement that is not possible with conventional pion beams. This is an example of how current electromagnetic pion production experiments are becoming sensitive to the final state interactions.
At this point the data for Re$E_{0+}$ below $k_{T}(\pi^{+}n)$ is not shown because it is very sensitive to the energy calibration of the tagger which is in progress. The preliminary data analysis does show the unitary cusp. The parameter $\beta$ can also be measured in the $\gamma \vec{p} \rightarrow \pi^{0}p$ reaction with a transverse polarized target. This experiment has been performed and the preliminary values of the data for the polarized cross section are shown in Fig.\ref{fig:E0p}\cite{A2-prop}. The data analysis is not yet complete so no value for $\beta$ can be presented at this time. This observable will allow us to extract the imaginary part of the electric dipole amplitude Im$E_{0+}$ and is potentially a more accurate way to determine $\beta$. When these experiments are completed they will provide a stringent test of ChPT calculations\cite{Cesar-fit,loop,Hilt} including isospin breaking in $\pi$N scattering lengths\cite{a_piN}. 

Compared to photo-pion production electro-pion production is in its early stages. The original experiments had shown a disagreement with HBChPT calculations\cite{BKM-electro}. A new experiment at Mainz corrected errors in the previously published values\cite{Merkel}, and these problems have now been resolved. The low energy constants of HBChPT\cite{BKM-electro} were fitted to the Mainz data by C. Smith and myself, and reasonable agreement was obtained. In addition the relativistic ChPT calculations\cite{Hilt} are also in agreement with these data\cite{Merkel}. There is also a JLab experiment carried out with the BigBite spectrometer which extends the kinematic range of the Mainz experiment, and a preliminary data analysis was presented[Lindgren]. In the region in which they overlap the two data sets agree. Electroproduction data has two relevant variables, the total energy W and the photon virtuality $Q^{2}$. It will be of interest to determine the maximum values of W and $Q^{2}$ for which the ChPT calculations agree with the data.

\section{Conclusions} 
In recent years we have seen substantial progress in confinement scale probes of QCD. These include experiment, ChPT, and lattice calculations. Space does not allow a full listing of all of the progress so I will only enumerate a few. There has been excellent progress in $\pi\pi$ scattering with the precise determination of the s wave $\pi\pi$ scattering lengths\cite{pi-pi,NA48}[Bizzeti]. Lattice calculations have confirmed the low energy constants used in ChPT\cite{lattice-ChPT}[Lellouch], validating that they closely approximate QCD. For the $\pi$N s wave scattering lengths there also has been excellent experimental\cite{Gotta} and theoretical\cite{a_piN} progress. There is good agreement between the results of pionic hydrogen and deuterium if the isospin breaking due to the mass difference of the up and down quarks is taken into account\cite{a_piN}. We have reached a new era in the prediction and experimental determination of the $\pi^{0}$ lifetime\cite{Goity,PrimEx,RMP}, although more needs to be done to bring the experiments up to the level of accuracy of the ChPT calculation. Progress has also been impressive in the $\gamma \pi$N sector, both in experiment\cite{Gotta} and in ChPT calculations\cite{a_piN}[Meissner] which take isospin breaking into account. Some ChPT calculations have started to include dispersion relations to take account of the neglected higher order terms\cite{Martin-dispersion,Lutz} as has been done in $\pi\pi$ scattering\cite{pi-pi}. Lattice calculations have reached the level of accuracy where they are strongly complementing the ChPT calculations\cite{lattice-ChPT}[Lellouch], and we anticipate even more progress in testing confinement scale QCD in the near future. I look forward to progress in all of this in the next few years and the next Chiral Dynamics workshop in 2015 in Pisa.

\end{document}